\documentclass[aps,prd,showkeys,showpacs,amssymb,cite,
amsfonts,epsf,preprintnumbers,nofootinbib,superscriptaddress]{revtex4}

\usepackage[dvips]{graphicx}
\usepackage{bm,latexsym,amsmath,amssymb,amsfonts,color}
\usepackage[mathscr]{eucal}
\newcommand{\be}[1]{\begin{equation} \label{#1}}
\newcommand{\ee}{\end{equation}}
\newcommand{\bea}{\begin{eqnarray}}
\newcommand{\eea}{\end{eqnarray}}
\newcommand{\ba}{\begin{array}}
\newcommand{\ea}{\end{array}}
\newcommand{\bel}{\begin{align}}
\newcommand{\eel}{\end{align}}
\newcommand{\nn}{\nonumber}
\newcommand{\tcb}{\textcolor{blue}}
\newcommand{\tcr}{\textcolor{red}}

\begin{document}

\title{Classifying Self-gravitating Radiations}
\author{Hyeong-Chan Kim}
\email{hckim@ut.ac.kr}
\affiliation{Department of Physics, North Carolina State University, Raleigh, NC 27695-8202, USA}
\affiliation{School of Liberal Arts and Sciences, Korea National University of Transportation, Chungju 380-702, Korea}%------------------------------------------------------------
%------------------------------------------------------------
%
\begin{abstract}
We study a static system of self-gravitating radiations confined in a sphere by using numerical and analytical calculations.
%We classify and analyze the solutions systematically.
Due to the scaling symmetry of radiations, most of main properties of a solution can be represented as a segment of a solution curve on a plane of two-dimensional scale invariant variables. 
We define an `approximate horizon' (AH) from the analogy with an apparent horizon.
Any solution curve contains a unique point which corresponds to the AH.
A given solution is uniquely labelled by three parameters representing the solution curve, the size of the AH, and the sphere size, which are an alternative of the data at the outer boundary.  
Various geometrical properties including the existence of an AH and the behaviors around the center can be identified from the parameters. 
We additionally present an analytic solution of the radiations on the verge of forming a blackhole.
Analytic formulae for the central mass of the naked singularity are given.
\end{abstract}
\maketitle
%
%\pagecolor[rgb]{0.90,1.00,0.90}
%\pagecolor{black}\color{white}{
\section{Introduction}
A bounded self-gravitating isothermal sphere are an interesting object as a model of a small dense nucleus of stellar systems, which is to some extent independent of the outer parts of the system~\cite{Lynden-Bell}.   
In 1981, a spherically symmetric solution of self-gravitating radiation was presented by Sorkin, Wald, and Jiu~\cite{Sorkin:1981wd} as a solution that maximizes the entropy based on general relativity.
The heat capacity and stability of the solution were further analyzed in Refs.~\cite{Pavon1988} and~\cite{Chavanis:2007kn,Chavanis:2001hd,Chavanis:2001ib}. 
Schmidt and Homann~\cite{Schmidt:1999tr} named the geometry `photon star'. 
Thereafter, the system has drawn attentions repeatedly in relation to the entropy bound~\cite{Schiffer:1989et,Hod:1999as}, maximum entropy principle~\cite{Gao:2016trd,Fang:2013oka,Gao:2011hh}, holograpic principle~\cite{Bousso:2002ju,Lemos:2007ys,Anastopoulos:2013xdk}, blackhole thermodynamics~\cite{Sorkin:1997ja}, and exclusion of blackhole firewalls~\cite{Page:2013mqa}. 
Self-gravitating radiation in Anti-de Sitter spacetime was also pursued~\cite{Page:1985em,Vaganov:2007at,Gentle:2011kv}.
Studies on the systems of self-gravitating perfect fluids are undergone~\cite{Pesci:2006sb}. 
An interesting extension of the self-gravitating system was presented in Ref.~\cite{Schmidt:1999tr,Anastopoulos:2011av} where a conical singularity is included at the center as an independent source from the radiation. 
The singularity was said to be benign in the sense that it does not give rise to inextensible causal geodesics.
Because no timelike geodesics reach the singularity and null geodesics simply pass it.
A thermodynamic interpretation for the conical singularity was also discussed~\cite{Anastopoulos:2011av}.
Analytic analysis was presented for an extreme case, which was interpreted as a spacetime blackhole with perfect fluid in equilibrium~\cite{Zurek}.  
Some of the singular solutions were shown to have an interesting geometry, which resembles an event horizon to an outside observer.
The similarity was used to understand the near horizon geometry of a blackhole~\cite{Anastopoulos:2014zqa}.

Let the sphere of radiations has a radius $R$.
Outside the sphere, the spacetime is described by the vacuum Schwarzschild metric with Arnowitt-Deser-Misner (ADM) mass $M_R$. 
The system is different from an ordinary star whose boundary is implemented by its own gravity and equation of state. 
In the absence of an artificial boundary at $R$, the density of the static solution becomes proportional to $1/r^2$ as $r\to \infty$. 
Then, the energy of the system goes to infinity and stability issues arise. 
Despite with this difference, we call it a `star' later in this work for convenience. 
The metric inside was shown~\cite{Anastopoulos:2011av} to be given by
\begin{equation} \label{metric}
ds^2 = - \left(1-\frac{2M_R}{R}\right)
	\sqrt{\frac{\rho(R)}{\rho(r)}} dt^2 + \frac{dr^2}{1-2m(r)/r} +r^2 
	(d\theta^2 + \sin^2\theta d\phi^2), \quad r \leq R,
\end{equation}
where $m(r)$ and $\rho(r)$ are the mass inside and the density at $r$, respectively.  
The energy density is related with the mass as 
\be{rhoP:m}
\rho(r) = \frac{1}{ 4\pi r^2 }\frac{dm(r)}{dr}.
\ee
Because the star is composed of radiations, it also satisfies $\rho(r) = 3 p(r)$, where $p(r)$ denotes the pressure at $r$. 
From the blackbody radiation law, the surface energy density is given by $\rho(R)= \sigma T^4$, where $T \equiv (1-2M_R/R)^{-1/2}\beta^{-1}$ denotes the locally measured temperature at $R$ and $\sigma$ is the Stefan-Boltzmann constant.  
In summary, a solution of the spherically-symmetric self-gravitating radiation can be uniquely specified by using the boundary data, 
\be{B}
\mathfrak{B} \equiv(R, M_R, T).
\ee

The structure of the star is described by the  Tollman-Oppenheimer-Volkoff (TOV) equation for the metric~\eqref{metric}.
Introducing a dimensionless variable 
\be{u:r}
u \equiv \frac{2m(r)}{r},
\ee
The TOV equation can be cast into a second order differential equation for $u$ as
\be{series}
\frac16 (u+u')(u+u'+3) + (u-1)\Big(u+u' -\frac{u'+u''}4 \Big) =0,
\ee
where prime denote the derivative with respect to a dimensionless variable 
\be{xi}
\xi = \log \frac{r}{r_H},
\ee
and $r_H$ is a length scale which will be specified later. 
The equation~\eqref{series} does not contain any length scale.
Therefore, the system possesses a scaling symmetry: The transformation $(m \to e^s m,~ r \to e^s r)$ preserves the equation of motions. 
As a result, various properties of the solutions can be described by scale invariant variables.
Introducing a scale-invariant variable $v$, 
\be{v:u}
v \equiv \frac{dm(r)}{dr} = 4\pi r^2 \rho(r) = \frac{u+u'}2,
\ee
the TOV equation~\eqref{series} can be rewritten as a first order differential equation for $u$ and $v$, 
\begin{equation} \label{dvdu}
\frac{dv}{du} = f(u,v) \equiv \frac{2v(1-2u-2v/3)}{(1-u)(2v-u)}.
\end{equation}

The integral curve of Eq.~\eqref{dvdu} on the $(u,v)$ plane is called a solution curve which is denoted by $C$ in this work. 
The allowed range of $(u,v)$ is $u<1$ and $v\geq 0$, where each inequality represents the fact that the spacetimes is static and the energy density of radiation is non-negative, respectively. 
For this reasons, we call $u=1$ and $v=0$ as the static boundary (SB) and the positive energy boundary (PB), respectively.  
From Eq.~\eqref{dvdu}, one may easily notice that any solution curve becomes parallel to the $u$ axis when it crosses PB and the line 
\be{P}
P: 2u+2v/3=1
\ee
and parallel to the $v$ axis when it crosses SB and the line
\be{H}
H: u = 2v.
\ee
Therefore, the solution curve cannot cross the two boundaries, SB and PB. 
This property also indicates that a solution curve will spiral around the point $(u,v)= (3/7,3/14)$, which is an end of every solution curves.
The solution curves are divided into two classes by means of the other end, which corresponds to the center of the star. 
One class is composed of a solution curve beginning at $(u,v) = (0,0)$.
The solution curve, namely $C$, describing solutions of pure radiations were found in Ref.~\cite{Sorkin:1981wd} as a stable static configuration, where the word `pure' implies that the star does not contain mass contributions other than the radiation.
The other class of solution curves describe spacetimes bearing a negative mass source at the center which presents a conical singularity.
The curves originate from $(u,v) \to (-\infty, 0)$.

%\tcr{Extend this part to include what is new in this work, what is the original result in this work?}\\
However, we notice that the understanding on the details of the geometry is still incomplete because its geometric structure is not classified systematically and analyzed throughly. 
The purpose of the present work is to make up for this part.
For the manuscript be self-consistent, we have taken some of previous results from literatures, which are in subSecs.~\ref{sec:scale}, \ref{sec:IIB}, \ref{sec:asym}, subSec. III A, and part of Sec.~\ref{sec:4}. 
One of the main purposes of this work is to characterize the spherically symmetric star of self-gravitating radiations by means of parameters representing their physical characteristics.
A solution corresponding to the second class were known to have a wall (deformed horizon-like object) which was named as `approximate horizon' (AH) in Ref.~\cite{Anastopoulos:2014zqa}.
As a first step to accomplish the purpose, in Sec.~\ref{sec:IID}, we present a clear definition for the AH from the analogy with an apparent horizon. 
Following the definition, each solution curve has a point corresponding to the AH irrespective of the existence of central singularity.  
As a second step, in Sec.~\ref{sec:IIE}, we classify the solution curves in terms of a physical parameter $\nu$ defined at the AH, where the value of $\nu$ is given by the distance of the solution curve $C_\nu$ from the SB.
Because the set of all possible solution curves sweeps every physically acceptable points in the $(u,v)$ plane, any point on the plane can equivalently be named by using $\nu$ and a parameter $\xi$ representing the position on the curve.  
Now, a solution of self-gravitating radiation is uniquely determined by the scale of the system, which is fixed by  the radius of the AH, $r_H$. 
A solution is uniquely described by using the boundary data in Eq.~\eqref{B}.  
Even though intuitive for an outside observer, this is inconvenient to describe the internal structure of the self-gravitating radiations contrary to the the description by using $(\nu,r_H, R\equiv r_H e^\xi)$.
 
We also present the general behaviors of solution curves including the singular one in Sec.~\ref{sec:IIIB}. 
Solution curves for various $\nu$ are presented and their behaviors are explained.
Specifically, the behaviors of solutions  are given as functions of $\nu$ around various positions including the center, the maximal $v$ position, and the AH.  
This compliments the shortage of the numerical calculation to find a solution curve. 
In Sec.~\ref{sec:4},  we provide an analytic solution describing the star on the verge of forming a blackhole composed of self-gravitating radiations.  
This is a generalization of the analytic analysis done in Refs.~\cite{Zurek,Anastopoulos:2014zqa}.
We summarize and discuss the results in Sec.~V.  

%=================
\section{Properties for self-gravitating radiation}
%=============================
\subsection{Scale transformation} \label{sec:scale}
Let us parametrize a solution curve $C_\nu$ of Eq.~\eqref{dvdu} in terms of $\xi$ as $(u(\xi), v(\xi))$, where $\nu$ is a parameter characterizing the curve which will be specified in subSec.~\ref{sec:IIE}.
The parametrization is determined by Eq.~\eqref{v:u} which presents the radius, $r(\xi)$. 
Note that, from Eq.~\eqref{xi}, the values of $u$ and $v$ at each point on the curve are invariant under the scaling,
\be{mr:scale}
m \to \bar m= e^s m, \quad r\to \bar r= e^s r, \quad \xi \to \bar \xi=\xi +s, 
\quad u(\xi) \to \bar u(\bar \xi) = u(\xi), \quad v(\xi) \to \bar v(\bar \xi) = v(\xi).
\ee
Therefore, $\bar u(\xi) = u(\xi-s)$ represents a new solution of Eq.~\eqref{series} with its mass, density, and radius are scaled.
Explicitly, the surface values $(u_R,v_R)$ are invariant,
\be{P:scale}
\bar u(\xi_R+s) = u(\xi_R) =\frac{2M_R}{R}, \quad 
\bar v(\xi_R+s) = v(\xi_R) =4 \pi R^2 \rho(R),
\ee
where $M_R = m(R)$.
On the other hand, the mass, the radius, and the density scale as 
\be{rP:scale}
\bar m(e^s R) = e^s M_R, \quad
\bar R = e^s R, \quad 
\bar \rho(e^s R)=e^{-2s} \rho(R) .
\ee
In addition, the range of $\xi$ is changed from $(-\infty, \xi_R]$ to $(-\infty, \xi_R+s]$. 
Given a solution $u(\xi)$ supported with the boundary value at $r=R$, 
the scale transformation provides a set of solutions having scaled mass and density related by Eq.~\eqref{rP:scale}. 
In fact, any point on $C_\nu$ can play the role of a boundary point which supports a set of solutions related by the scale transform.
In this sense, a solution curve $C_\nu$ provides a set of solutions parameterized by $R$ and $s$.  

%======================
\subsection{The central region} \label{sec:IIB}
Let us study the behaviors of the solutions to Eq.~\eqref{series} around the origin with $\xi \ll -1$.
Eq.~\eqref{series} allows two different small $r$ behaviors,
\begin{align} 
m_{\rm r}(r) = \frac{\kappa r_H}{6} \left(\frac{r}{r_H}\right)^3 ,  
\qquad
m_{\rm s}(r) = \frac{r_H}{2}\left[-\mu_0 +\frac{\kappa}{5} \left(\frac{r}{r_H}\right)^5
\right] . \label{sol:sing}
\end{align}
We scale the mass and radius in terms of $r_H$ to encompass the scale symmetry. 
The first one, $m_{\rm r}(r)$, corresponds to the regular solution which will be discussed in subSec.~\ref{sec:reg}. 
The second one, $m_{\rm s}$, must be negative at the origin, implying $\mu_0>0$.
Because the radiation has positive energy density, $\kappa$ is required to be positive.
The energy density of radiation at the center is 
$\rho_{\rm s} = (\kappa/8\pi r_H^2) \times (r/r_H)^2 $.
Comparing the functional forms of $\rho_{\rm s}$ with $\rho_{\rm r}$, it is evident that the radiation is repelled from the center by the negative mass at the central singularity.

We next consider how the solutions~\eqref{sol:sing} change with respect to the scaling. 
Through the scale transformation~\eqref{mr:scale}, the coefficients $\kappa$ and $\mu_0$ should change.
Let us consider the case of regular solution, $m_{\rm r}(r)$, first.
Under the scaling in Eq.~\eqref{mr:scale}, the functional form of $m_{\rm r}$ in Eq.~\eqref{sol:sing} changes to 
\be{barmr}
\bar m_{\rm r}(\bar r) = e^s m_{\rm r}(r) \quad \Rightarrow \quad
\bar m_{\rm r}( \bar r) = \frac{\kappa \bar r_H}{6} \left(\frac{\bar r}{\bar r_H}\right)^3,
\ee
where we use the transformation law of the referential point~\eqref{rP:scale} and $\bar r_H = e^s r_H$ is the transformed radius of $r_H$.
Omitting bar in $\bar{r}$ in Eq.~\eqref{barmr}, the functional form of the scaled solution is the same as the original one with the replacement $\kappa \to \kappa \,e^{-2s} $.
Therefore, once one finds a solution for given $\kappa$, the central form of all other solutions related by the scaling is also at hand.

For the case of the solution $m_{\rm s}(r)$, the scaling~\eqref{mr:scale} leads to 
\be{barms}
\bar m_{\rm s}(\bar r) =\frac{\bar r_H}{2} \bigg[- \mu_0  + \frac{\kappa}5 \left(\frac{\bar r}{\bar r_H}\right)^5 \bigg].
\ee
Note that the functional form of the scaled solution is, omitting bar in $\bar{r}$, the same as the original one with the replacements $ \mu_0 \to e^s \mu_0$ and $\kappa \to e^{-4s} \kappa$.
With this form, the explicit functional dependence on the scale becomes evident. 
Note that the mass, $-\bar r_H \mu_0$, of the conical singularity changes with the scale.

%======================
\subsection{Asymptotic behaviors} \label{sec:asym}
We next consider the asymptotic behavior as $\xi\to \infty$.
By assuming solutions in the form, $ u = a e^{-\xi} + b$, one may find that the differential equation~\eqref{series} allows two different asymptotic forms for $u(\xi)$,
\be{vh:asym}
u_a \rightarrow \frac37, \qquad u_t \rightarrow a e^{-\xi}
,
\ee
where $a$ is an arbitrary constant.
With the increase of $\xi$, $u_a$ acts as if it is an attractor which pulls every nearby solutions (Here, we interpret as if $\xi$ is a time even though it is a radial coordinate). 
Introducing a small perturbation, the solution of Eq.~\eqref{series} with first nonvanishing contribution takes the  form,
\be{ua1}
u_a =\frac37\Big[1+ \tilde c_1 e^{-3\xi/4} 
	\cos\Big(\frac{\sqrt{47}}{4} \xi +\tilde\phi \Big)\Big] \quad
	 \Rightarrow  \quad
m_a(r) = \frac{3r}{14}\Big[1+\tilde c_1\Big(\frac{r_H}{r}\Big)^{3/4} 
	\cos\Big(\frac{\sqrt{47}}{4} \log \frac{r}{r_H} +\tilde \phi \Big) \Big],
\ee
where $\tilde c_1$ and $\tilde \phi$ are integration constants.
On the $(u,v)$ plane, the curves satisfies
\be{uv:asym}
\Big(u-\frac{3}7\Big)^2 + \frac{64}{47} \Big(v-\frac38 u-\frac3{56}\Big)^2 
	= \Big(\frac{3 \tilde c_1}{7}\Big)^2 \Big( \frac{r_H}{r}\Big)^{3/2}.
\ee
The solution curves spire in to the point $(u,v)=(3/7,3/14)$ as $r$ increases. 
For the asymptotic form~\eqref{ua1}, the scale transform~\eqref{mr:scale} is achieved by $r_H \to \bar r_H$.
%\tcr{One may fix the value of $\tilde c_1$ once we set $r= r_H$.}

The other asymptotic form, $u_t(\xi)$, approaches zero and $v_t(\xi) =0$ for all $\xi$.
The asymptotic form plays the role of an attractor when the radius $r$ (equivalently $\xi$) decreases from infinity.
Small perturbations around $u_t$
\be{ut}
u =u_t+\delta(\xi) =  a e^{-\xi} + \epsilon  \,e^{2\xi} 
\quad \Rightarrow \quad m_t(r) = \frac{a}{2} + \frac{\epsilon}{2} \left(\frac{r}{r_H}\right)^3,
\ee 
increase indefinitely as $r\to \infty$ and the solution leaves $u_t$ permanently,
where $\epsilon$ is a tiny integration constant. 
However, as will be shown later, in the presence of an AH close to an apparent horizon, 
this almost constant mass region becomes wide enough to regard Eq.~\eqref{ut} as an approximate asymptotic form for $r \gg r_H$. 

Eq.~\eqref{vh:asym} may not be realized unless the outer boundary is located in the asymptotic region.
For most cases, noting the heat capacity in Ref.~\cite{Pavon1988}\footnote{The heat capacity was calculated for a regular solution. 
However, the same formula holds for other solutions up to the contribution of central singularity.}, an instability appears at a smaller radius than $r_H$ which makes the star unstable. 
However, as noted in Ref.~\cite{Anastopoulos:2011av}, the conical singularity may play some role in thermodynamics and affects to the stability.  
In this sense, the issue of stability needs additional research.

%=====================
\subsection{Approximate horizon } \label{sec:IID}
Solutions other than the regular one have a conical singularity at the center~\cite{Anastopoulos:2011av}.
Based on Newtonian gravity, objects having positive masses attract each other. 
Similarly, the radiations are attracted to the center due to their self-gravity. 
On the other hand, the negative mass at the center repels the radiations. 
The balance of the two behaviors concentrates the radiations at some intermediate region.
Therefore, in the presence of large negative mass at the center and a large quantity of radiations outside enough to compensate the repulsion, a region of strong gravity will be developed.
To represent this phenomena, we define an `approximate horizon' (AH).

Consider a metric of the form,
$$
ds^2 = h_{ab}(x^c) dx^a dx^b + r^2 d\Omega_{(2)},
$$
where $a,b,c = 0,1$ and $r(x^a)$ is a scalar function of $x^0$ and $x^1$.
An apparent horizon for the metric is defined by the surface satisfying 
\be{AH}
\chi^2 \equiv h^{ab} \nabla_a r \nabla_b r =0.
\ee 
At the apparent horizon, the vector field $\nabla_a r$ becomes null.  
For the metric ansatz~\eqref{metric}, the coordinate change $dr \to 0$ for a finite proper distance change $\delta r = \sqrt{g_{rr}} dr$.
Applying Eq.~\eqref{AH} to the metric~\eqref{metric}, the apparent horizon will exist only at the position satisfying $\chi^2  = 1- 2m(r)/r =0$.
Equation~\eqref{dvdu} indicates that this condition is never achieved with the self-gravitating radiation because $du=0$ at the SB, which implies that a solution curve never cross the SB.

If $\chi^2$ is very small at a surface surrounded by thermal radiations, an observer located outside of the region has difficulty in distinguishing the surface from an apparent horizon. 
Even though photons may propagate over the surface, it would be difficult to discern it from the surroundings.  
In this sense, we define an AH as a surface where $\chi^2 $ takes its minimum value for a given solution curve.
The local minimization condition of $\chi^2$ for the metric~\eqref{metric} becomes
\be{chi}
 \delta\big(\chi^2 \big) =0 \quad
\Longrightarrow \quad 
 u = 2v, \qquad  \frac37 < u \leq 1,
\ee
where the inequality is included because $\chi^2$ is locally maximized on $H$ for $u<3/7$.  
Noting the asymptotic form~\eqref{uv:asym}, a solution curve crosses the line $H$ many times. 
Of all the points, the nearest one to the SB corresponds to the AH. 
In fact, as can be seen in the next section, every AH are located in the region $ u_r \leq u < 1$, where $u_r \approx 0.4926$ corresponds to the AH value of regular solution.

%======================================
\subsection{Characterizing solution curves} \label{sec:IIE}
 
\begin{figure}[th]
\begin{center}
\begin{tabular}{cc}
\includegraphics[width=.4\linewidth,origin=tl]{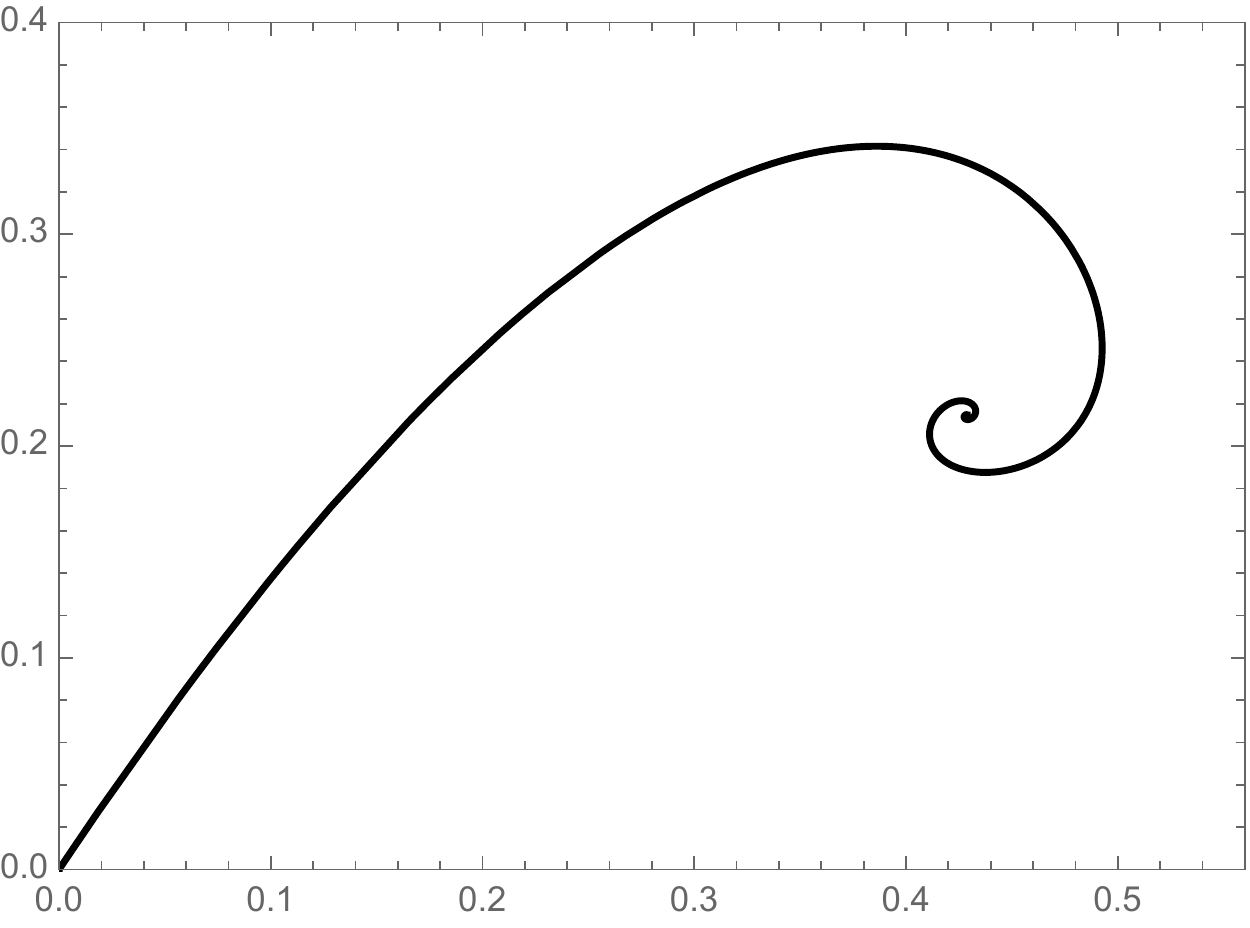}&
\qquad
\includegraphics[width=.4\linewidth,origin=tl]{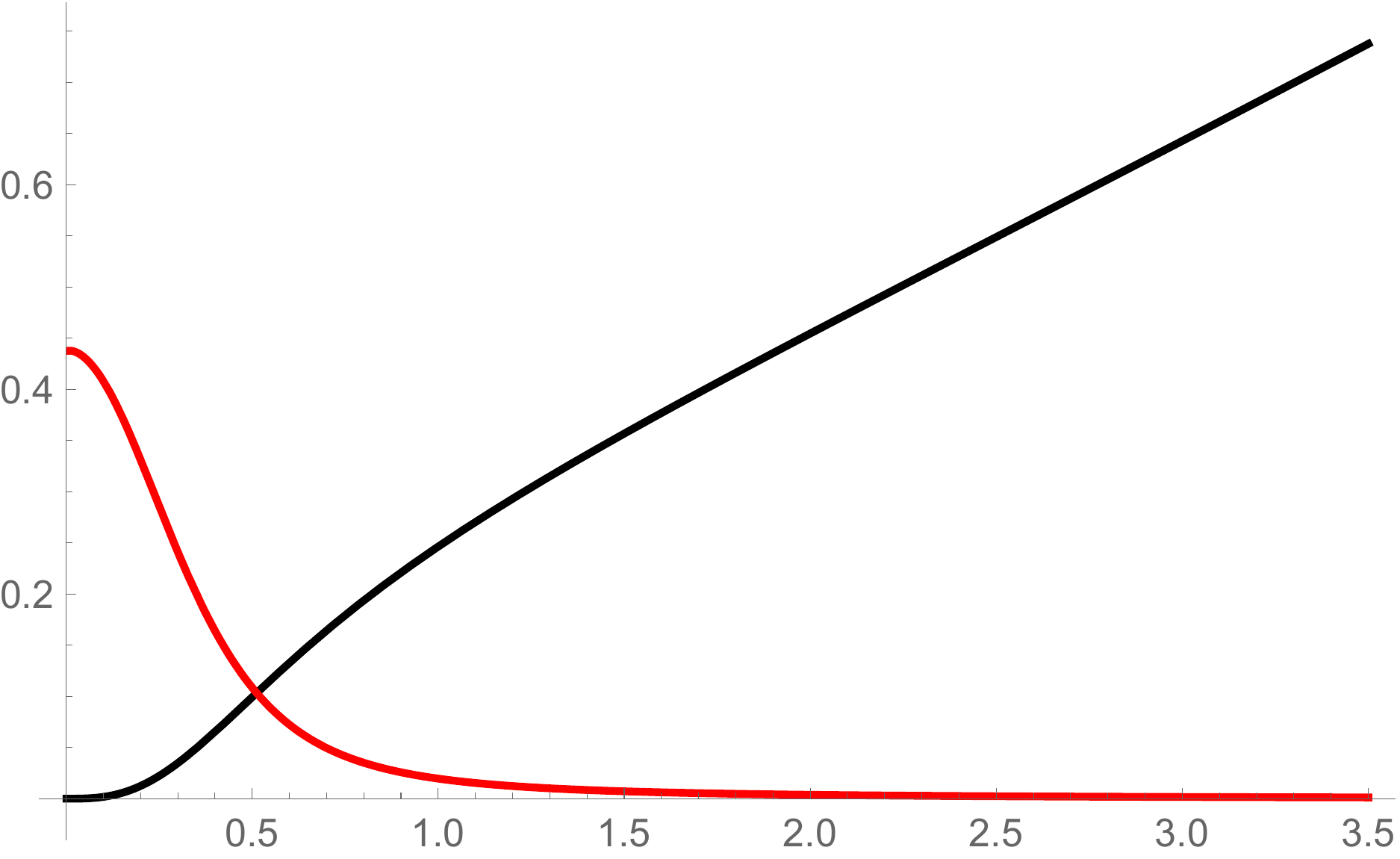}
%(a) Solution curve $C$ & (b) The mass and density profile
\end{tabular}
\put (-20,-75) {$r/r_H$  }
\put (-435,65) { $v$}
\put (-243,-70) {$u$  }
\put (-429,-59) {$O$} \put (-425.5,-64.5){$\bullet$}
\put (-291,60) {$P$} \put (-291,53.5){$\bullet$}
\put (-274.5,-7) {$P'$} \put (-274.5,0){$\bullet$}
\put (-262,20.5) {$H$} \put (-255,20.5){$\bullet$}
\put (-272,9) {$\mathcal{R}$} \put (-277,9){\tcb{$\bullet$}}
\put(-65,40) {$m(r)/M$}
\put(-50,-55) {\tcr{$ M^2 \rho(r)$}}
\end{center}
\caption{The solution curve and typical form for the mass and density profile for a regular solution. 
The solution curve in the left panel reproduces Fig.~1 in Ref.~\cite{Anastopoulos:2011av}.
%(b) shows typical forms of the mass and density profile. 
Each point on $C$ corresponds to a boundary data for a regular solution.
}
\label{fig:reg}
\end{figure}
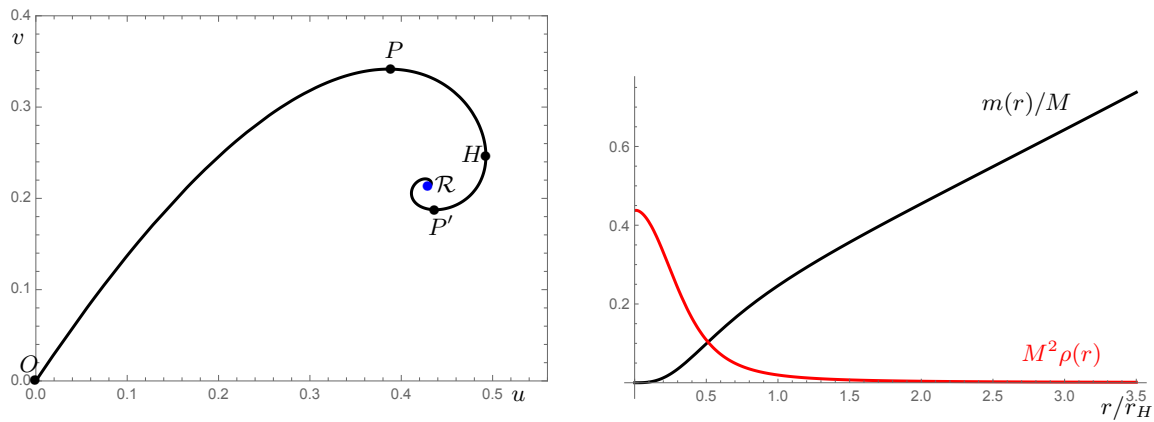

In Refs.~\cite{Sorkin:1981wd,Chavanis:2007kn,Anastopoulos:2011av}, 
the authors identified the solution space in terms of a set of parameters defined at the surface $r=R$.
To do this, Eq.~\eqref{series} was integrated to obtain the solution curve $C$ inward to $r=0$ after fixing the boundary value $(u_R,v_R)$.
A lesson from the experiences on the regular solution is that a solution space is conveniently characterized by the combination of a point $(u_R,v_R)$ on the solution curve and the radius $R$ of the system as in Fig.~\ref{fig:reg}.
The whole solution space can be found once one finds a set of solution curves covering whole physically acceptable region of $(u,v)$ plane. 
Because the $(u,v)$ plane is two dimensional, the one-dimensional solution curves should be characterized by one parameter, namely, $\nu$. 
An immediate task is to find out a most convenient parameter for $\nu$.
A first impression is to use the central values such as $\mu_0$ or $\kappa$ in Eq.~\eqref{sol:sing}. 
For example, with the choice of a scaling $e^s= \kappa^{1/4}$, one can choose $\nu \equiv \bar\mu_0 = \mu_0 \kappa^{1/4}$ with $\bar \kappa =1$.
This choice is relevant theoretically only for singular solutions and the behaviors of the solution is too sensitive to a small change of the central values numerically.

At the present work, we choose to characterize the solution curve in terms of the distance of the solution curve to the SB:
\be{nu}
\nu \equiv 1-u_H.
\ee 
%at the AH. 
Because any solution curve spires in to the point $\mathcal{R}$, it meets the line $H$ indefinite number of times. 
To avoid multiple labeling, we should restrict the value of $\nu$ so that any solution curve has only one name. 
This can be achieved by restricting the value of $\nu$ so that its maximum value corresponds to that of the regular solution $\nu_r  \approx 0.50735$, which value can be determined from the results in Ref.~\cite{Sorkin:1981wd}. 
Now, the line $H$, which labelling $C_\nu$, is given by
\be{BC2}
H \equiv \big\{H_\nu(1-\nu, \frac{1-\nu}2 )|~ 0 < \nu \leq \nu_r  \big\}, 
\ee
where the absence of horizon restricts $\nu > 0$. 

For a given $C_\nu$, the value of $u$ is maximized to be $u_H=1-\nu$ at $H_\nu$.
By using the scale transform, we may freely choose a referential solution so that $\xi=0$ at $H$.
Let us consider a system given by the boundary data $(R, M_R, T_R)$. 
The boundary values for the scale invariant variables are given by $u_R = 2M_R/R$ and $v_R = 4\pi \sigma R^2T_R^4$.
The solution curve can be parametrized by $\xi = \log (r/r_H)$. 
Then, the position of the boundary data $(u_R,v_R)$ on the solution curve can be uniquely represented once $\xi_R = \log(R/r_H)$ is known. 
Therefore, a specific solution is uniquely determined once we know
\be{solution}
(\nu, r_H, R).
\ee
Now, it is easy to identify whether or not there is an AH. 
If $\xi_R $ is positive, the star size $R$ is larger than $r_H$ and a AH exists.
Under the scale transformation, $(\nu, r_H, R)$ is mapped to $(\nu, e^s r_H, e^s R)$. 
$\xi_R$ is invariant because it is defined from the ratio of two radius of the star and of the AH, both are co-variant.

%==========================
\section{Various solutions} %
Any point $(u_R,v_R)$ on the solution curve $C_\nu$ can play the role of a boundary data at $r=R$.
Because of the scale invariance~\eqref{P:scale}, a given point $(u_R,v_R)$ represents a one parameter family of boundary data,
$\{ (e^s M_R, e^s R, e^{-2s} T), s\in \Re \}$, characterized by the scaling $s$.   
Rather than using the boundary data, we characterize a solution by identifying $(\nu, r_H,R)$.

%=======================
\subsection{Regular solutions} \label{sec:reg}
Before dealing general solutions, let us illustrate a well-known solutions in Refs.~\cite{Sorkin:1981wd,Pavon1988,Chavanis:2007kn}. 
Most results in this subsection are reproductions of theirs except a few details.
The purpose of this subsection is to illustrate the role of a solution curve $C$ in $(u,v)$ plane obtained by numerical integration of Eq.~\eqref{dvdu} or Eq.~\eqref{series}.
The curve $C$ is given in the left panel of Fig.~\ref{fig:reg}.
An important property of the solution curve is its uniqueness.
The value $\nu_r $ presented just after Eq.~\eqref{BC2} is given by obtaining $u_H \approx 0.4926$ for the solution curve.
A typical mass and density profiles for a specific solution are shown in Fig.~\ref{fig:reg} (b). 

The solution curve $C$ begins with $u=0=v$ at $r=0$.
On the curve  $v$ takes its maximal value $v_P \approx 0.3416$ at $P$ and 
$u$ takes its maximum $u_H$ at $H$.
The ratio of the radial coordinates between $P$ and $H$ is $r_P/r_H \equiv e^{\xi_P} \approx 0.4823$. 
The allowed range of $u$ and $v$ are restricted to $0\leq u \leq u_H$ and $0 \leq v \leq v_P$, respectively. 
Eventually, the curve spires in to the point $\mathcal{R}(3/7,3/14)$.  
From the analysis of the heat capacity in Ref.~\cite{Pavon1988}, a thermodynamic instability is set on for a system having the outer boundary in the region from $P$ to $H$.
Interestingly, the heat capacity for a system having the outer boundary just outside the point $H$ is positive definite.

The density at the center is $\rho(0)= \kappa (8\pi r_H^2)^{-1}$, where $\kappa \approx 4.759$, and monotonically decreases with $r$. 
The central form of mass is given by $m_{\rm r}(r)$ in Eq.~\eqref{sol:sing}. 
For $r \gg r_H$, the mass increases linearly.

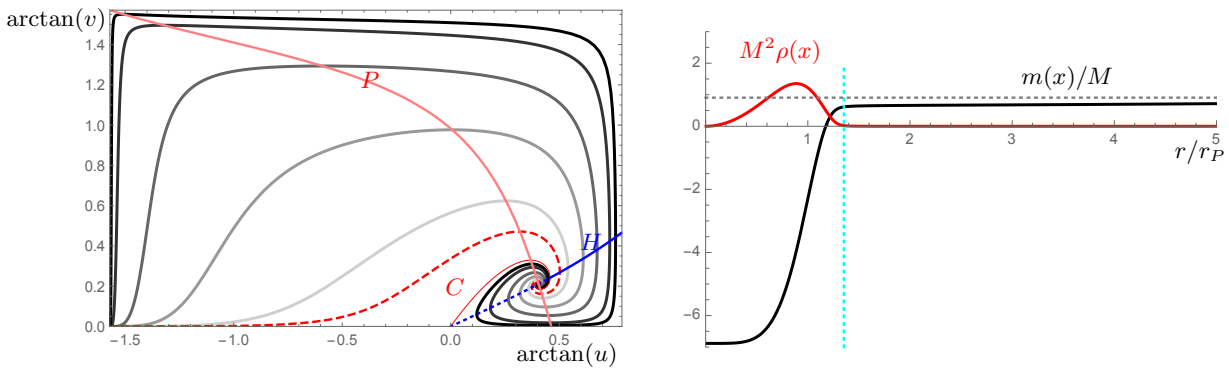
\begin{figure}[tb]
\begin{center}
\begin{tabular}{cc}
\includegraphics[width=.4\linewidth,origin=tl]{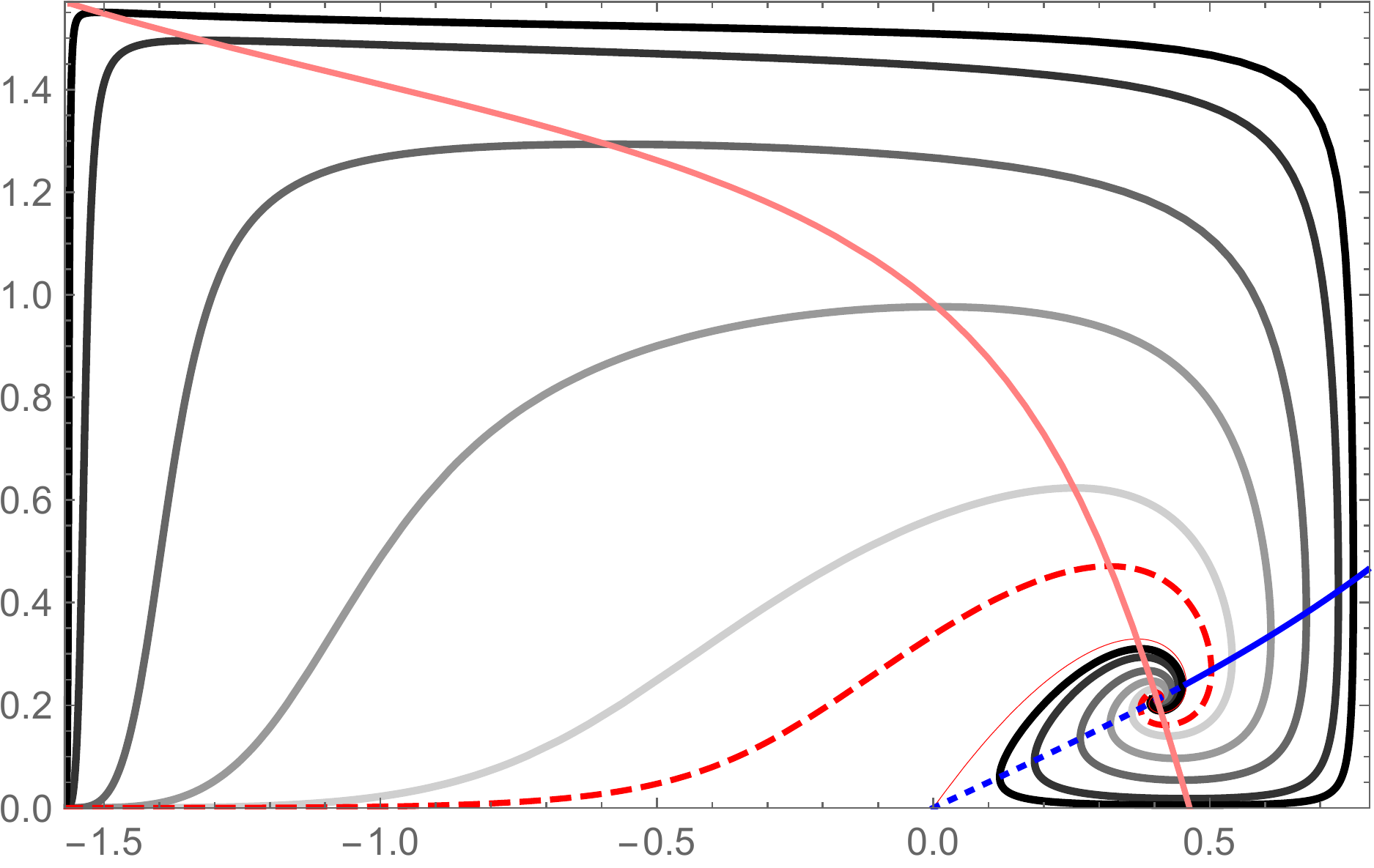}&
\qquad
\includegraphics[width=.4\linewidth,origin=tl]{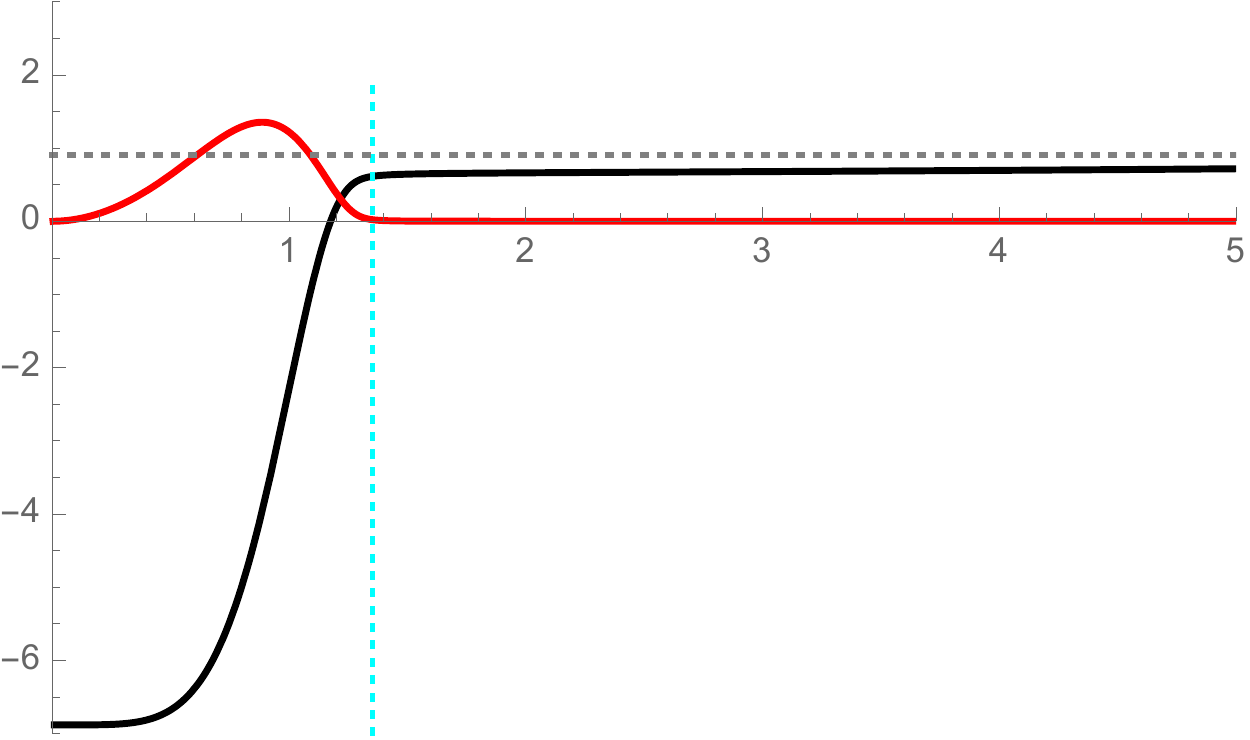}
%\\
\end{tabular}
\put (-185,50) {$\tcr{M^2 \rho (x)}$  }
\put (-20,12) {$r/r_{P}$  }
\put (-245, -23) {\tcb{$H$}}
\put (-465,62) { $\arctan(v)$}
\put (-269,-65) {$\arctan(u)$  }
\put (-78,40) {$m(x)/M$}
\put (-328,38.5){\tcr{$P$}}
\put (-299,-40){ \tcr{$C$}}
\end{center}
\caption{The solution curves (Left), mass and density distribution (Right). 
In the left panel, each curve corresponds to $\nu=\nu_r, 0.45, 0.4 , 0.3, 0.2, 0.1,$ and $0.05$, respectively from the bottom.  
The blue curve $H$ represents the referential position given in Eq.~\eqref{H}.
The thick pink curve $P$ denotes the position where $v$ is maximized on solution curves in Eq.~\eqref{P}.
In the right panel, the typical behavior of mass and density is given. 
This figure corresponds to $\nu=.1$. 
The dotted cyan line denotes where a AH forms.
}
\label{fig:fh}
\end{figure}
%====================
\subsection{Solutions with a conical singularity at the center  }
	\label{sec:IIIB}
In the left panel of Fig.~\ref{fig:fh}, we plot solution curves for several different values of $\nu$ to show their characteristic behaviors. 
In the right panel, the typical behaviors of the mass and the density for $\nu < \nu_r$ with respect to the radius are additionally displayed. 
As in Eq.~\eqref{nu}, $\nu$ represents the closest distance from the SB to the solution curve $C_\nu$.
Therefore, along $H$, a solution curve with a smaller $\nu$ are nearer to the SB than that with a larger one.
The solution curve $C \equiv C_{\nu_r}$ representing the regular solution is the farthest one from the SB in weak red form. 
A solution curve $C_\nu$ with $0< \nu < \nu_r$ is located outside of it and the corresponding solution has a conical singularity at the center. 
The radius $r$ increases as one tracking a given solution curve $C_\nu$ in $\mathcal{R}$ in Fig.~\ref{fig:reg}. 
This result comes from Eq.~\eqref{v:u}, which gives $d\xi = du/(2v-u)$.
From this one notice that the radius of the sphere in Eq.~\eqref{xi} grows when $u$ increases above $H$ and decreases below $H$.
Because the gradient of $u$ changes sign on $H$, the radius monotonically increases if one tracing the solution curve inward.
In addition, the solution curve spirals in to the point $\mathcal{R}$ as $r\to \infty$ because it should be horizontal (vertical) on $P$ ($H$) and should not cross itself.

\begin{figure}[hbt]
\begin{center}
\begin{tabular}{cc}
\includegraphics[width=.4\linewidth,origin=tl]{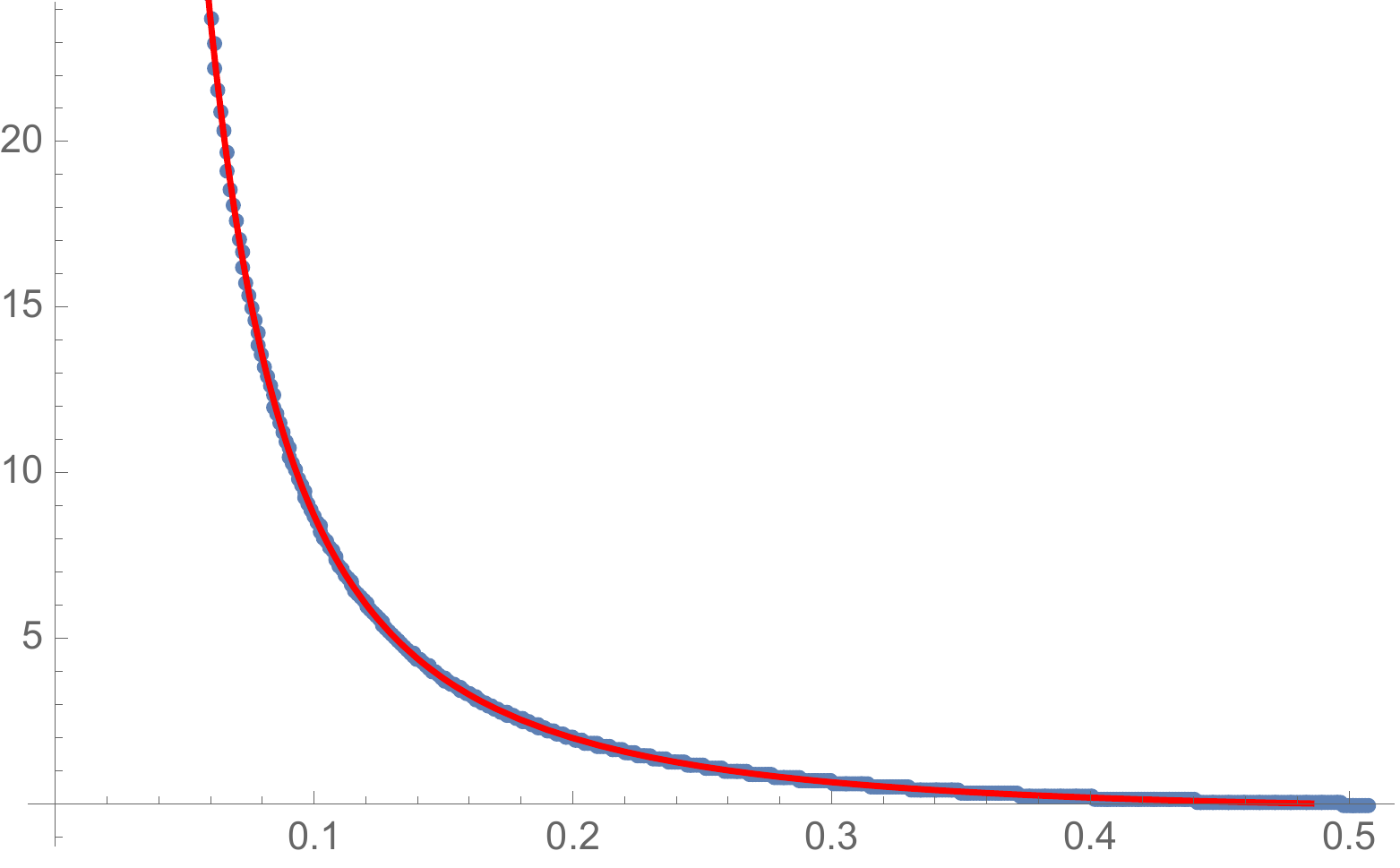}&
\qquad
\includegraphics[width=.4\linewidth,origin=tl]{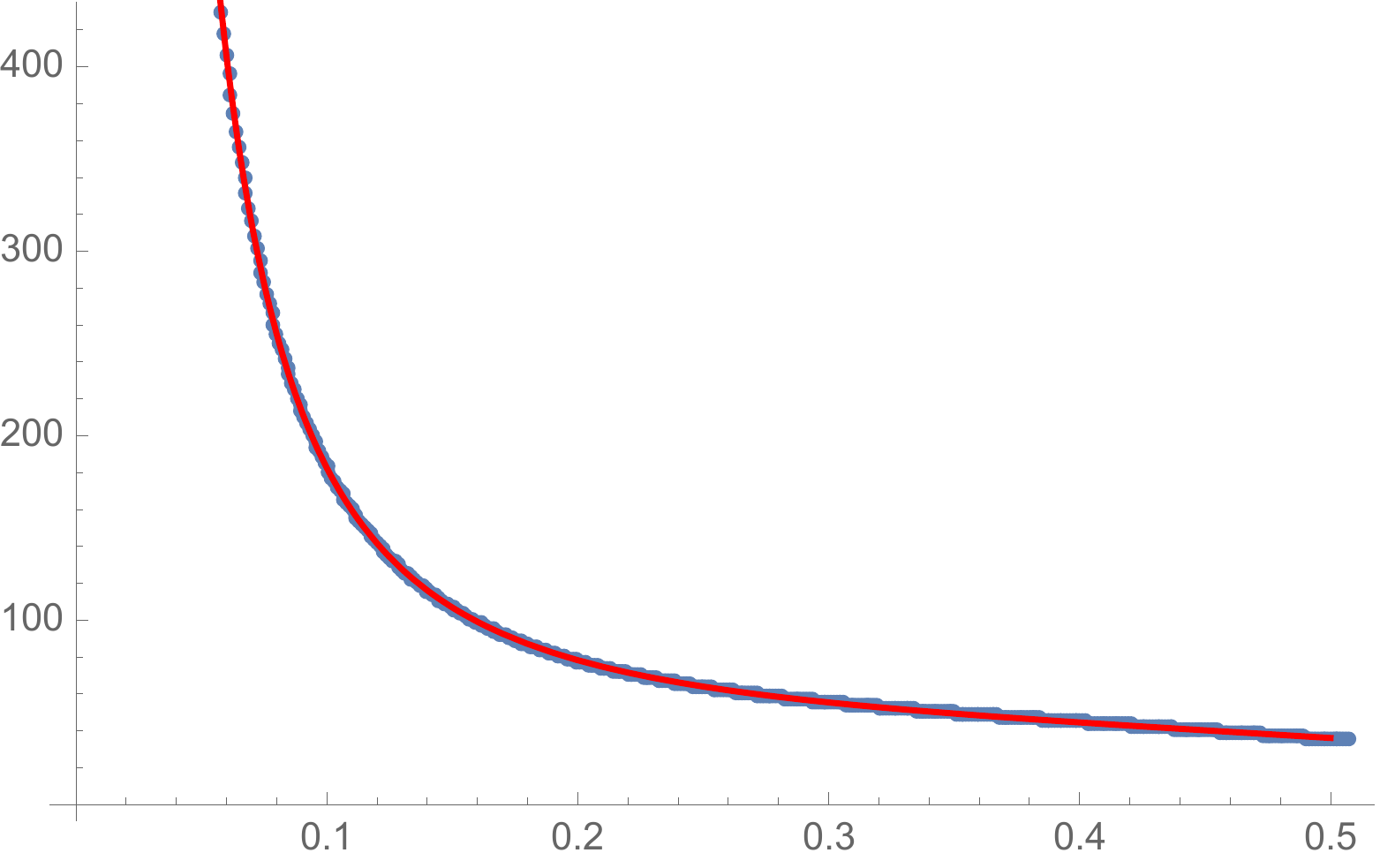}
\\
(a) $\mu_0$ as a function of $\nu$. & (b)  $\kappa$ as a function of $\nu$.
\end{tabular}
\put (-205,50) {$\kappa$  }
\put (-245, -53) {$\nu$}
\put (-435,65) { $ \mu_0$}
\end{center}
\caption{The change of $\mu_0$ and $\kappa$ as a function of $\nu$. 
The approximate functional form are given as red curves.
}
\label{fig:mukappa}
\end{figure}

Differences of $C_\nu$ from the regular-solution curve $C$ are apparent for small $r$.
The curves begin at $(u,v) \to (-\infty,0)$ at $r=0$, where the negative value of $u$ is due to the negative mass of conical singularity, $m(0) = -\mu_0 r_H/2$. 
In addition, the density of the radiations vanishes at the center and increases quadratically with $r$. 
Comparing the result of regular solution where the density takes maximum value at the center, the repulsive nature of the central negative mass to positive energy radiation is apparent.
From Eqs.~\eqref{u:r}, \eqref{v:u}, and \eqref{barms}, one may find that $v u^4\approx \kappa \mu_0^4/2 $ for $r  \sim 0$.
Therefore, a solution curve $C_{\nu}$ stays close to the PB for $u \ll 0$. 
The size of $\kappa \mu_0^4/2$ determines how fast will the solution curve depart the PB. 
An interesting inspection is that both $\kappa$ and $\mu_0$ are not independent variables but are functions determined by the value of $\nu$ only, which functions are numerically plotted in Fig.~\ref{fig:mukappa}.
Both monotonically decreases with $\nu$ and diverge inverse quadratically as $\nu \to 0$.
Numerically, the graphs in Fig.~\ref{fig:mukappa} are well approximated by
\bea \label{mukappa}
 \mu_0 &\approx & 0.0771 \nu^{-2} + 0.2136 \nu^{-1} -1.2077 + 1.0067 \nu+\cdots, \\
 \kappa &\approx & 1.049 \nu^{-2}+ 5.872 \nu^{-1}+ 12.93 +
	71.90 \nu -114.7 \nu^2+ \cdots, \nn
\eea
where the range is $0< \nu < \nu_r$. 
A discrepancy of this numerical result is that the plots in Fig.~\ref{fig:mukappa} fails to reproduce the regular solution result for $\nu= \nu_r$ where $\mu_0 =0$ and $\kappa \approx 4.759$.
This is because the equations in Eq.~\eqref{mukappa} are expanded around $\nu =0$.
Let us consider the case with $\nu \sim 0$ briefly.   
Because $\kappa \mu_0^4 \propto \nu^{-10} \gg 1$, the value $v$ of a solution curve can be nontrivially large even for large negative $u$.
The size of $v$ is comparable to that of $|u|$ when they are of $O(\nu^{-2})$.  

\begin{figure}[th]
\begin{center}
\begin{tabular}{cc}
\includegraphics[width=.4\linewidth,origin=tl]{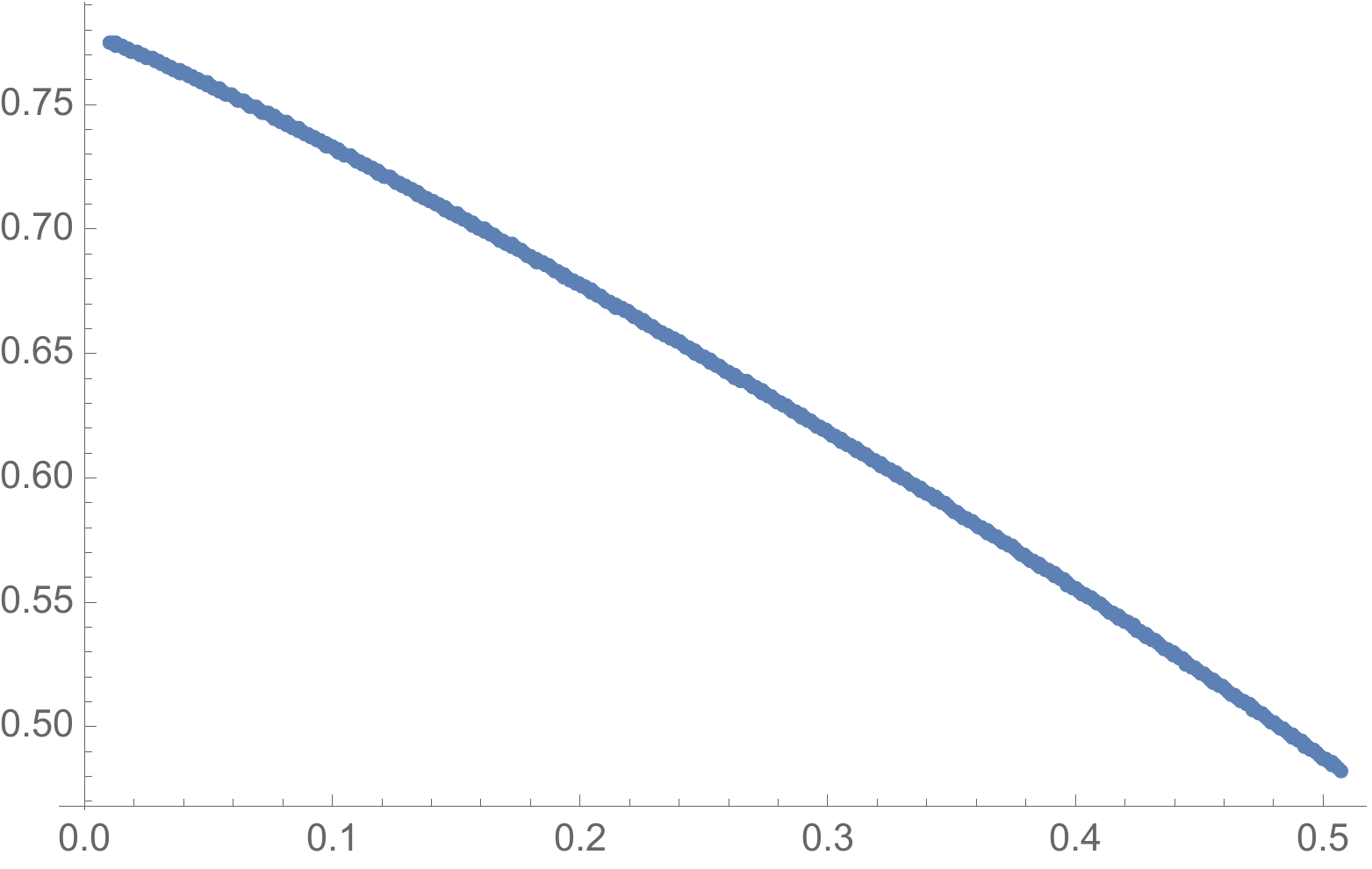}
~&~
%\qquad
\includegraphics[width=.4\linewidth,origin=tl]{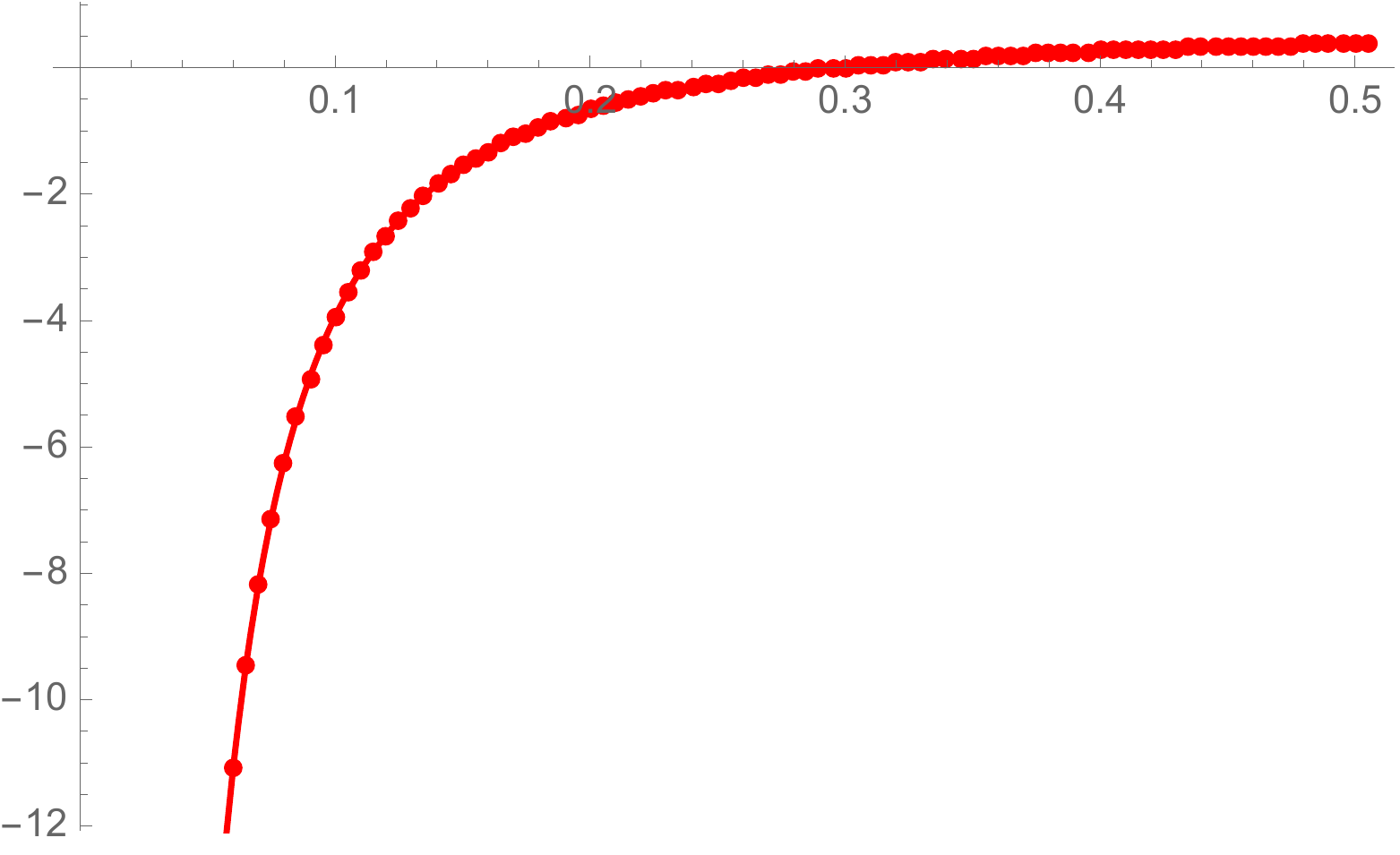}
\end{tabular}
\put (-423,62) {$\frac{r_{P}}{r_H}$} 
\put(-238,-60) {$ \nu$}
\put (-210,-50) {$u_P$} 
\put(-30,60) {$ \nu$}
\end{center}
\caption{The values of $r_P/r_H$ and $u_P$ with respect $\nu$. 
}
\label{fig:xiP}
\end{figure}

As the radius increases, the value of $v$ will be maximized at the point $P_\nu$ where $C_\nu$ crosses $P$. 
The maximum value of $v= v_{P_\nu}$ monotonically increases as $\nu$ decreases from $\nu_r$.
The value of $u_{P_\nu} = 1/2-v_{P_\nu}/3$ at the maximum point is plotted in the right panel of Fig.~\ref{fig:xiP}. 
The graph can be approximated to be
$$
u_{P_\nu} \approx-0.03558\nu^{-2}-0.1271\nu^{-1}+0.9246-0.3858 \nu + \cdots.
$$ 
For $\nu \ll 1$, it behaves inverse quadratically as expected in the previous paragraph.  
The value $r_{P_\nu}/r_H$ decreases with $\nu$ as in the left panel of Fig.~\ref{fig:xiP}.
The function can be approximated to be 
\be{P/H}
\frac{r_{P}}{r_H} = 0.7807 -0.4519 \nu - 0.3383 \nu^2 +\cdots .
\ee
%The ratio monotonically decreases with $\nu$.  
At $\nu=\nu_r$, the ratio becomes $r_P/r_H \approx 0.4823$, which value is the same as that of the regular solution. 
Note that the ratio and $u_{P_\nu}$ are functions of $\nu$ only, i.e., they are independent of the boundary choice and the scale.
For $u> u_P$, the value of $v$ decreases until the solution curve meets the line $P$ once more. 

As the radius further increases, the value of $u$ is maximized at the point $H_\nu$ where $C_\nu$ crosses the line $H$. 
The maximum value of $u$ and the corresponding value for $v$ is given by
\be{uv:H}
u_{H} = 1-\nu, \qquad v_H = \frac{1-\nu}{2}.
\ee
Around the point $H_\nu$, small $\nu $ behavior of $C_\nu$ is interesting. 
The density gradient becomes large because $dv/d\xi \propto (1-u)^{-1} \approx \nu^{-1}$.  
On the other hand, the mass gradient may not be large because $du/d\xi = du/dv \cdot dv/d\xi$ is $O(1)$. 
A detailed analysis for the behavior of the solution curve will be displayed in the next section.

When the solution curve is located below the line $H$, the value of $u$ starts to decrease with $r$ until $C_\nu$ meet $H$ again. 
Outside the point, $C_\nu$  converges gradually to that of the regular solution as it spires in to the point $\mathcal{R}$ in Fig.~\ref{fig:reg}.
The asymptotic approach is described by Eq.~\eqref{uv:asym}.

%=====================
\section{Analytic treatment of a self-gravitating radiation on the verge of forming a blackhole}
	\label{sec:4}
The geometry around an AH with $\nu \sim 0$ is an especially interesting object because it is on the verge of becoming a true event horizon in the sense that $m(r_{\rm fh}) \approx r_{\rm fh}/2$. 
We hope that this interesting case allows an analytic description, which turns out to be true.
Part of the present results were given in Refs.~\cite{Zurek,Anastopoulos:2014zqa}.
In Ref.~\cite{Zurek}, the authors assumed the temperature to be the Hawking temperature.
At the present work, we do not take the assumption because it is not always proper. 
An AH is different from a blackhole horizon even though it has some similarity.
In Ref.~\cite{Anastopoulos:2014zqa}, the author analyzed the near AH geometry analytically. 
Equations \eqref{1-u} and \eqref{v:u2} overlaps with Eqs.~(71) and (66) in Ref.~\cite{Anastopoulos:2014zqa}, respectively.
We extend their analysis to include the whole range of the system. 
%Equations~\eqref{1-u} and \eqref{v:u2} in this section overlaps with Eqs. (71) and (66) in the reference, respectively. 

\begin{figure}[bht]
\begin{center}
\begin{tabular}{cc}
\includegraphics[width=.4\linewidth,origin=tl]{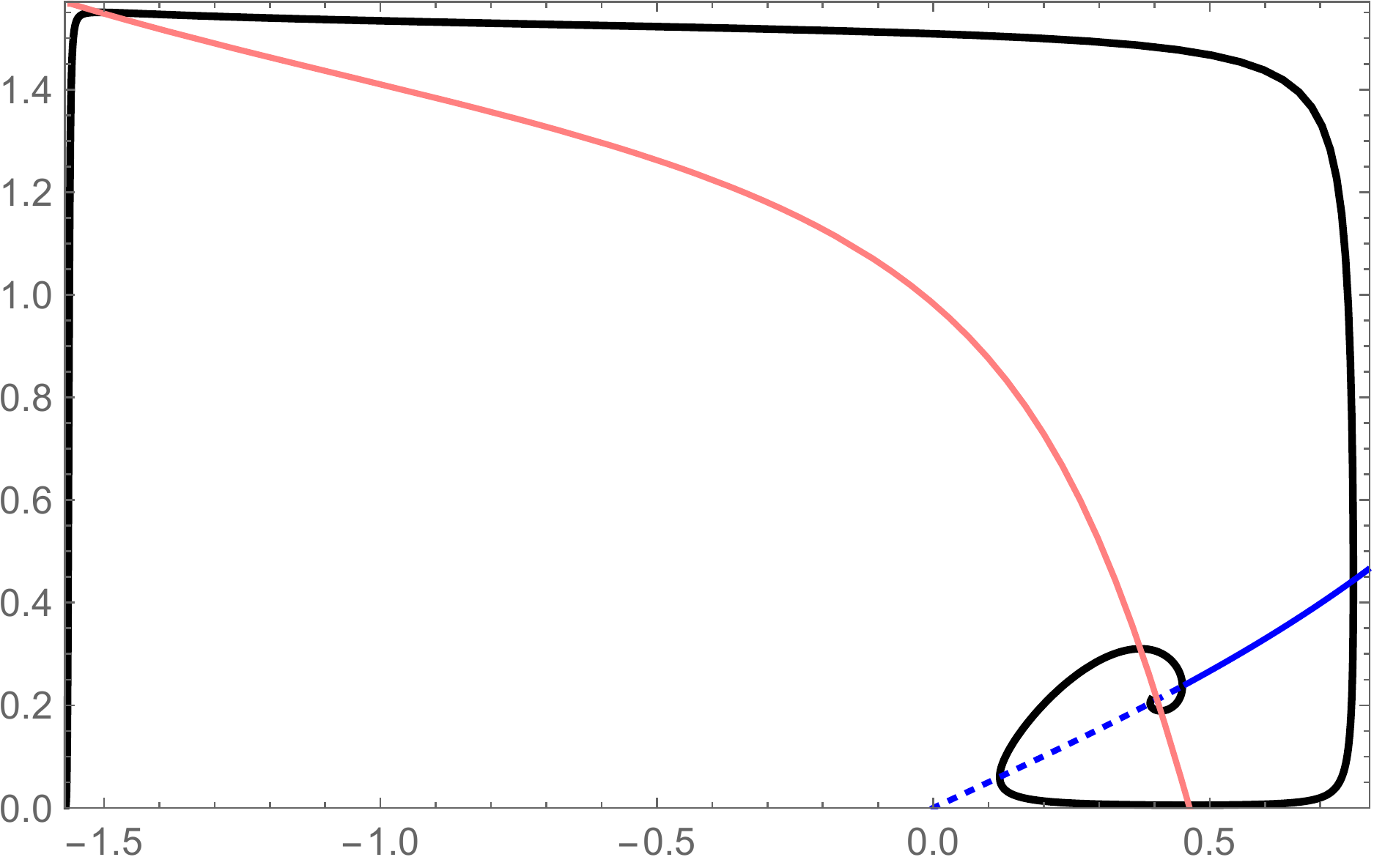}
%&\qquad \includegraphics[width=.4\linewidth,origin=tl]{rmi2m.pdf}
%\\
%(a) 
\end{tabular}
%\put (-207+100,78) {$\chi(x)=x-m(x)/M$  }
%\put (-10+100,-57) {$x$  }
\put (-197,-49) { $\mathfrak{I} $} \put (-198,-53) {$\bullet $}
\put (-198,57) { $P_\nu$} \put (-193,64) {$\bullet $}
\put (-31,-48) { $P'_\nu$} \put (-31.5,-53){$\bullet $}
\put (-1,-21) {$H_\nu$  } \put (-7,-19.5) {$\bullet $}
%\put (-30,50) {$m(x)/M$}
\put(-17, 29) {$\mathfrak{R}$} \put (-8,29) {$ \bullet$}
\put(-15, -48) {$\mathfrak{S}$} \put (-9,-50) {$ \bullet$}
\put(-68, -46) {$\mathfrak{S'}$} \put (-60,-49.5) {$ \bullet$}
\put (-238,62) { $\arctan(v)$}
\put (-20,-59) {$\arctan(u)$  }
\put(-100,30) {\tcr{$P$}}
%\textcolor{cyan}{
%\put(-415,50){Newtonian}}
\end{center}
\caption{ Solution curve (black) for systems having strong AH. 
The point $\mathfrak{I}$ represents $(u,v) = (-\infty,0)$. 
}
\label{fig:approx}
\end{figure}
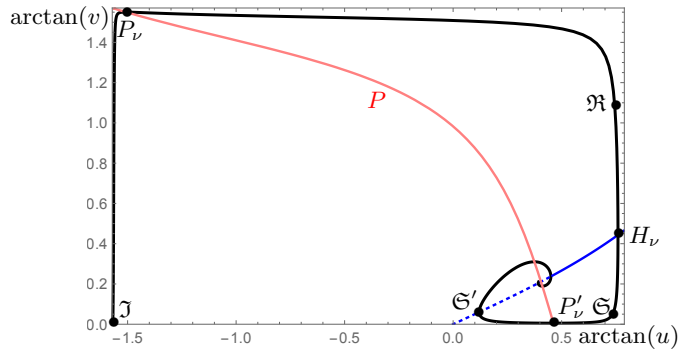
A solution curve $C_\nu$ is given in Fig.~\ref{fig:approx}.
For simplicity, we represent a segment of the solution curve by using its boundary points. 
For example, $[ P_\nu, \mathfrak{ R}]$ represents the segment from the point $P_\nu$ to the point $\mathfrak{R}$.
An AH is in the segment $[\mathfrak{R},\mathfrak{S}]$.
There, $u$ is almost constant but $v$ changes a lot. 
The segment $[\mathfrak{S,S}']$ is located just outside of the AH.
$v$ remains close to zero and $u$ decreases monotonically with $r$.
For $\nu \approx 0$, most of the mass is located inside the AH.

First, let us analyze the segment $[\mathfrak{R},\mathfrak{S}]$ of the solution curve depicted in Fig.~\ref{fig:approx}, which denotes the region near AH. 
On the segment, $u$ changes slowly with $u \approx u_H \approx 1$.
%, which  plays the role of a parameter measuring how close the AH is to the event horizon.  
Solving Eq.~\eqref{dvdu} keeping first non-vanishing corrections of $O(\nu)$, one get 
\be{1-u}
1-u \approx  \varepsilon \frac{ (2v/3+1)^2}{\sqrt{2v}} + O(\varepsilon^2),
\qquad   \frac12 \varepsilon^2 \ll v_{\mathfrak{S}} <   v  < v_{\mathfrak{R}}\ll  \frac12 \Big(\frac{9}{\varepsilon}\Big)^{2/3},
\ee
where $ \varepsilon = 9\nu/16$ is the expansion parameter.
We choose the points $\mathfrak{R}$ and $\mathfrak{S}$ so that 
$v_{\mathfrak{R}} = \varepsilon^{-1/3}$ and $v_{\mathfrak{S}} = \varepsilon^{2/3}/2$.
Integrating Eq.~\eqref{v:u} and using Eq.~\eqref{xi}, we get 
\be{rv:Q}
r =r_H e^\xi ; \qquad \xi  =\frac{\varepsilon }{\sqrt{2v}}\left(1 -\frac{v}{6} \right)- \frac{11\varepsilon}{12},
\ee
where we choose $\xi =0$ at $H$ ($v=1/2$).
Note that $r$ changes only a bit for a large change of $v$ in $[\mathfrak{R},\mathfrak{S}]$.
The mass inside $r$ is, keeping to the dominant corrections, given by
\be{mass2}
m(r(v)) = \frac{r_H}2 e^{\xi} u \approx \frac{r_H}{2}\left[ 1
	-\frac{\varepsilon (2v)^{3/2}}{9} \Big(1+\frac{27}{8v}\Big) -\frac{11\varepsilon}{12}\right] .
\ee
The relative increase of the mass from $\mathfrak{R}$ to $\mathfrak{S}$ is 
$
(m_\mathfrak{S}-m_\mathfrak{R})/m_\mathfrak{R} \approx 2^{3/2}/9 \times \varepsilon^{1/2},
$
which is relatively small compared to $m_\mathfrak{R}$.
On the other hand, the density, $\rho = v/4\pi r^2$, decreases rapidly in $[\mathfrak{R},\mathfrak{S}]$ so that $\rho(r_{\mathfrak{S}})/\rho(r_{\mathfrak{R}}) \approx \varepsilon/2$, because $\rho$ is roughly proportional to $v$.
This implies that the density profile in $r$ is extremely steep.

We next analyze the segment $[\mathfrak{S},\mathfrak{S}']$ of the solution curve depicted in Fig.~\ref{fig:approx}, where $u_\mathfrak{S} \approx 1$ and $0<u_{\mathfrak{S}'} \ll 1$.
In the region of interest, $v \ll 1$ and $dv < du$.
The solution curve $C_\nu$, solving Eq.~\eqref{dvdu} up to first nonvanishing order, is described by
\be{v:u2}
v = \frac{\varepsilon^2}{2u^2(1-u)^2};
\qquad 
\varepsilon^{2/3} \ll u_{\mathfrak{S}'} 
	\equiv \varepsilon^\beta <  u < u_\mathfrak{S} = 1- \varepsilon^{2/3}.   
\ee
We identify the two curves in Eqs.~\eqref{1-u} and \eqref{v:u2} at $\mathfrak{S}$, satisfying $du=dv$.
The radius can be obtained by integrating Eq.~\eqref{v:u} using Eq.~\eqref{v:u2}. 
To the first nonvanishing correction, we get
\be{xi:S} 
r= r_H e^\xi  \approx \frac{r_H}{u} 
	\exp\left[-\frac{11\varepsilon}{12}+ \frac{\varepsilon^2}{3u^3}\right]
\quad \Rightarrow \quad
u \approx \frac{r_H }{r} \left[1-\frac{11\varepsilon}{12}+ \frac{\varepsilon^2}{3} \left(\frac{r}{r_H}\right)^3  \right],  
\ee 
where we keep terms to $O(\varepsilon^1)$ and we choose $\beta = 1/3$.
Now, the range of the radius for the segment is given by $ r_H< r< \varepsilon^{-1/3} r_H$.
The radius increases almost inverse linearly with $u$.
The segment $[\mathfrak{S},\mathfrak{S}']$ corresponds to the long quasi-asymptotic region given by $u_t(\xi)$ in Eq.~\eqref{ut}.
Note that the mass $m(r) = r_H/2$ is included in the surface $r= (11/4\varepsilon )^{1/3} r_H$.

Outside the surface $\mathfrak{S}'$, near $0< u,v\ll 1$, the solution curve approaches that of the regular solution.
The function $u$ takes exactly the same form as that in Eq.~\eqref{xi:S}.
On the other hand $v = (\varepsilon^2/2)\times (r/ r_H)^2$. 
Therefore, from the point of view of a far outside observer, the geometry appears to be a combination of a constant point mass, $M = r_H(1-11\varepsilon/12)/2$, surrounded by a constant density, $\rho_0 = \varepsilon^2/(8\pi r_H^2)$.  
This result determines a local temperature uniquely, 
$$
T  = \frac{1}{(8\pi \sigma)^{1/4}} \sqrt{\frac{\varepsilon}{r_H}}. 
$$

We next consider the segment $[\mathfrak{I},\mathfrak{R}]$ which corresponds to the region far inside the AH. 
In the region, $1-u \gg 0$ and $2v+1-u \gg 1$. 
As was done in Ref.~\cite{Zurek}, Eq.~\eqref{dvdu} and ~\eqref{v:u} can be approximately solved to give 
\be{v:z}
v \equiv  \frac{15/2}{(r_H/r)^5-1} (1-u) = \frac12 \Big(\frac{3}{5\varepsilon}\Big)^{2} \, 
	\Big(\frac{r}{r_H}\Big)^4 \left[1-\Big(\frac{r}{r_H}\Big)^5\right]^2.
\ee
$v$ takes its maximum value $v_{\rm max} = (7/2)^{1/5} 9/ 7^3\varepsilon^2 $ at $r = r_{\rm max} \equiv (2/7)^{1/5} r_H$, where the integration constant is determined by matching the value of $1-u$ at $\mathfrak{R}$ with Eqs.~\eqref{1-u} and~\eqref{rv:Q}.
The value of $u$ at the maximum point is $u_{\rm max} = 1-v_{\rm max}/3$.
From Eq.~\eqref{dvdu}, the point that maximize $v$ must be located on the line $P$ given in Eq.~\eqref{P}, which is not satisfied by $(u_{\rm max}, v_{\rm max})$. 
This indicates that Eq.~\eqref{v:z} is an approximate solution with error of $O(1/v_{\rm max})$, which becomes negligible if the limit $v_{\rm max} \gg 1$ is taken.  
Note also that $r_{\rm max}/r_H = (2/7)^{1/5} \approx 0.7783$ is close to the numerically fitting value in Eq.~\eqref{P/H}.

Finally, comparing the $r\to 0 $ limit of Eq.~\eqref{v:z} with Eqs.~\eqref{barms} then using Eqs.~\eqref{u:r} and \eqref{v:u}, we get the limiting forms for $\nu \ll 1$ as
\be{mu0kappa}
\mu_0 = \frac{2}{15} \Big(\frac{16}{15}\Big)^2 \nu^{-2}, \qquad 
\kappa =  \Big(\frac{16}{15}\Big)^2 \nu^{-2}.
\ee
This implies that both $\mu_0$ and $\kappa$ diverge quadratically as $\nu \to 0$.
Both coefficients are close to their numerical fitting values in Eq.~\eqref{mukappa}.
Note that the resulting form for the bare mass 
\be{m0}
m(0)\equiv -\frac{r_H\mu_0}2 \propto -\frac{r_H}{\nu^2} 
\ee
is formally different from that in Ref.~\cite{Zurek} ($\propto -r_H^3$). 
The difference comes from the fact that the temperature was treated to be that of the blackhole in the reference. 
The density at the center,
\be{rho0}
\rho(r) \to \left(\frac{16}{15}\right)^2\frac{1}{8\pi \nu^2} \frac{r^2}{r_H^4},
\ee 
also grows extremely fast.

%===================================
\section{Summary and discussions}
We studied a system of self-gravitating radiation confined in a spherical box by using numerical and analytic calculations.
We classified the solution space systematically and defined an `approximate horizon' (AH) from the analogy with an apparent horizon.  
We also analyzed an analytic solution describing a radiation star on the verge of forming a blackhole.
%having an AH whose geometry is close to an apparent horizon.

Assuming the box to be large enough so that its boundary is located in the asymptotic region, the behavior of the solutions can be summarized as follows: 
The geometry outside the box is described by the Schwarzschild metric.
Around the center, the mass behaves as $m_{\rm r} \propto r^3$ or $m_{\rm s} \propto -m_0 + r^5$ in the absence or in the presence of a central conical singularity, respectively, where $r$ is the areal radial coordinate.
On the other hand, for large $r$, the asymptotic form follows a single formula~$m_a\propto 3r/14$.
In between the two limits, an AH appears.
If the AH is close to an apparent horizon,
the transient state $m_t \sim r_{\rm fh}/2 + \epsilon^2 r^3$ arises for a wide range of $r$ just outside of the AH, where $\epsilon$ is a very small number.

To classify the solutions, we have designed a set of parameters which characterize both the internal geometry and the macroscopic quantities. 
The first parameter $\nu$ identifies a solution curve $C_\nu$, a curve on the two dimensional space of scale invariant variables $(u,v) \equiv (2m(r)/r, 4\pi r^2 \rho(r))$,  which satisfy a first order differential equation originated from the Tollman-Oppenheimer-Volkoff equation.
Here, $\nu \equiv 1-u_H$ represents the orthogonal distance from the solution curve $C_\nu$ to the line $u=1$, where the distance is measured at the AH.
In this sense, $\nu$ measures how much an AH differ from an apparent horizon. 
The value of $\nu$ is limited to a finite domain $0\leq \nu \leq \nu_r \approx 0.50735$.
For $\nu=0$, the AH becomes an apparent horizon. 
%Only for $\nu=\nu_r$, the solution is absent of central singularity.
Any solution other than the regular solution with $\nu=\nu_r$ has a conical singularity at the origin.
The second parameter $r_H$ represents the size of the AH. 
It determines the scale of the star described by the solution curve $C_\nu$.
Any point on $C_\nu$ is parameterized by $(u(\xi), v(\xi))$ where $\xi = \log (r/r_H)$ where $u(\xi)$ and $v(\xi)$ are parameterized functions of $\xi$.  
The last parameter is the star size $R$.  
Given the three parameters $(\nu, r_H, R)$, the ADM mass and the surface temperature are given by $M_R=  R u(\xi_R)/2$ and $T= (v(\xi_R)/ 4\pi \sigma R^2)^{1/4}$, respectively, where $\xi_R = \log(R/r_H)$. 
Many important physical properties are determined by the sign of $\xi_R$.
% position of the boundary on the solution curve.  
For example, if $\xi_R$ is positive ($R> r_H$), the box includes the AH.
If $\xi_R$ is negative ($R < r_H$), the system does not include an AH and the contribution of the radiation to the mass will be maximized around the surface.
For the case of a regular solution, the sign of the heat capacity for a self-gravitating system~\cite{Pavon1988} is also determined by the position of the boundary on the $(u,v)$ plane.

With respect to the behavior near the central singularity, 
we found a mass formulae, $m(0) \propto - r_H/\nu^2$, approximately with respect to the variation of $\nu$.
In the $\nu \to 0$ limit, the mass function diverges inverse quadratically which was shown to be correct by using numerical calculation.  
The density per unit area also diverges inverse quadratically.
The mass and density formula will be useful when one study the behavior of the central conical singularity with respect to the change of macroscopic quantities.  
In studying the stability of the star, the heat capacity plays a central role. 
However, the calculation of the heat capacity is nontrivial because of the central singularity of which thermodynamic properties we do not know. 
If it is possible to ignore the singularity, the heat capacity will be determined by the physical values at the boundary similarly to the case of the regular solution in Ref.~\cite{Pavon1988}.
In that case, the heat capacity for systems having boundary just inside of the AH is negative definite.
On the other hand, the heat capacity just outside is positive definite even with the geometrical similarity to the event horizon. 
Far outside of it, a wide approximately constant mass region appears.
The geometry in this region resembles that of the Schwarzschild blackhole with a radiation field.
Systems having boundary in this region have negative heat capacity once more. 
Whereas, there is a proposal for the thermodynamic properties of the central singularity~\cite{Anastopoulos:2011av}.
Because of this complexity, the stability issues~\cite{Chavanis:2001hd,Chavanis:2001ib}
should be dealt cautiously. 
A way to avoid the complexity due to the central singularity is to place an inner boundary after excising the central part. 
In this case, the solutions can be used to study a spherical shell of self-gravitating radiations. 

\section*{Acknowledgment}
This work was supported by the National Research Foundation of Korea grants funded by the Korea government NRF-2013R1A1A2006548.
%%%%%%%%%%%%%%%%%%%%%


\begin{thebibliography}{99}

\bibitem{Lynden-Bell}
D. Lynden-Bell and R. Wood, 
%``The Gravo-Thermal Catastrophe in Isothermal Spheres and The Onset of Red-Giant Structure for Stellar Systems",
Mon. Not. R. Astr. Soc. {\bf 138}, 495 (1968).

%\cite{Sorkin:1981wd}
\bibitem{Sorkin:1981wd} 
  R.~D.~Sorkin, R.~M.~Wald and Z.~J.~Zhang,
  %``Entropy of selfgravitating radiation,''
  Gen.\ Rel.\ Grav.\  {\bf 13}, 1127 (1981).
  %%CITATION = GRGVA,13,1127;%%
  %87 citations counted in INSPIRE as of 24 Apr 2015

\bibitem{Pavon1988}
D.~Pavon and P. T.~Landsberg, 
%``Heat Capacity of a Self-Gravitating Radiation Sphere,"
Gen.\ Rel.\ Grav.\ {\bf 20}, 457 (1988).

%\cite{Chavanis:2007kn}
\bibitem{Chavanis:2007kn} 
  P.~H.~Chavanis,
  %``Relativistic stars with a linear equation of state: analogy with classical isothermal spheres and black holes,''
  Astron.\ Astrophys.\  {\bf 483}, 673 (2008)
  [arXiv:0707.2292 [astro-ph]].
  %%CITATION = ARXIV:0707.2292;%%
  %15 citations counted in INSPIRE as of 24 Apr 2015
  
%\cite{Chavanis:2001hd,Chavanis:2001ib}
\bibitem{Chavanis:2001hd} 
  P.~H.~Chavanis,
  %``Gravitational instability of finite isothermal spheres,''
  Astron.\ Astrophys.\  {\bf 381}, 340 (2002)
  doi:10.1051/0004-6361:20011438
  [astro-ph/0103159].
  %%CITATION = doi:10.1051/0004-6361:20011438;%%
  %52 citations counted in INSPIRE as of 10 Dec 2015

%\cite{Chavanis:2001ib}
\bibitem{Chavanis:2001ib} 
  P.~H.~Chavanis, C.~Rosier and C.~Sire,
  %``Thermodynamics of selfgravitating systems,''
  Phys.\ Rev.\ E {\bf 66}, 036105 (2002)
  doi:10.1103/PhysRevE.66.036105
  [cond-mat/0107345].
  %%CITATION = doi:10.1103/PhysRevE.66.036105;%%
  %30 citations counted in INSPIRE as of 10 Dec 2015
  

%\cite{Schmidt:1999tr}
\bibitem{Schmidt:1999tr} 
  H.~J.~Schmidt and F.~Homann,
  %``Photon stars,''
  Gen.\ Rel.\ Grav.\  {\bf 32}, 919 (2000)
  [gr-qc/9903044].
  %%CITATION = GR-QC/9903044;%%
  %7 citations counted in INSPIRE as of 20 May 2015

%\cite{Schiffer:1989et}
\bibitem{Schiffer:1989et} 
  M.~Schiffer and J.~D.~Bekenstein,
  %``Proof of the Quantum Bound on Specific Entropy for Free Fields,''
  Phys.\ Rev.\ D {\bf 39}, 1109 (1989).
  doi:10.1103/PhysRevD.39.1109
  %%CITATION = doi:10.1103/PhysRevD.39.1109;%%
  %53 citations counted in INSPIRE as of 01 Dec 2015
  
%\cite{Hod:1999as}
\bibitem{Hod:1999as} 
  S.~Hod,
  %``Entropy bounds for rotating systems and the nine souls of the generalized second law,''
  gr-qc/9901035.
  %%CITATION = GR-QC/9901035;%%
  %13 citations counted in INSPIRE as of 01 Dec 2015
 
%\cite{Gao:2016trd}
\bibitem{Gao:2016trd} 
  S.~Gao,
  %``A General Maximum Entropy Principle for Self-Gravitating Perfect Fluid,''
  Springer Proc.\ Phys.  {\bf 170}, 359 (2016).
  doi:10.1007/978-3-319-20046-043
  %%CITATION = doi:10.1007/978-3-319-20046-0_43;%%  
  
%\cite{Gao:2011hh}
\bibitem{Gao:2011hh} 
  S.~Gao,
  %``A general maximum entropy principle for self-gravitating perfect fluid,''
  Phys.\ Rev.\ D {\bf 84}, 104023 (2011)
  [Phys.\ Rev.\ D {\bf 85}, 027503 (2012)]
  doi:10.1103/PhysRevD.84.104023, 10.1103/PhysRevD.85.027503
  [arXiv:1109.2804 [gr-qc]].
  %%CITATION = doi:10.1103/PhysRevD.84.104023, 10.1103/PhysRevD.85.027503;%%
  %15 citations counted in INSPIRE as of 01 Dec 2015  

%\cite{Fang:2013oka}
\bibitem{Fang:2013oka} 
  X.~Fang and S.~Gao,
  %``General proof of the entropy principle for self-gravitating fluid in static spacetimes,''
  Phys.\ Rev.\ D {\bf 90}, no. 4, 044013 (2014)
  doi:10.1103/PhysRevD.90.044013
  [arXiv:1311.6899 [gr-qc]].
  %%CITATION = doi:10.1103/PhysRevD.90.044013;%%
  %7 citations counted in INSPIRE as of 01 Dec 2015
  
%\cite{Lemos:2007ys}
\bibitem{Lemos:2007ys} 
  J.~P.~S.~Lemos,
  %``Black hole entropy and the holographic principle,''
  arXiv:0712.3945 [gr-qc].
  %%CITATION = ARXIV:0712.3945;%%
  %2 citations counted in INSPIRE as of 01 Dec 2015


%\cite{Anastopoulos:2013xdk}
\bibitem{Anastopoulos:2013xdk} 
  C.~Anastopoulos and N.~Savvidou,
  %``The thermodynamics of self-gravitating systems in equilibrium is holographic,''
  Class.\ Quant.\ Grav.\  {\bf 31}, 055003 (2014)
  doi:10.1088/0264-9381/31/5/055003
  [arXiv:1302.4407 [gr-qc]].
  %%CITATION = doi:10.1088/0264-9381/31/5/055003;%%
  %5 citations counted in INSPIRE as of 01 Dec 2015  

%\cite{Bousso:2002ju}
\bibitem{Bousso:2002ju} 
  R.~Bousso,
  %``The Holographic principle,''
  Rev.\ Mod.\ Phys.\  {\bf 74}, 825 (2002)
  doi:10.1103/RevModPhys.74.825
  [hep-th/0203101].
  %%CITATION = doi:10.1103/RevModPhys.74.825;%%
  %703 citations counted in INSPIRE as of 01 Dec 2015  
  
%\cite{Sorkin:1997ja}
\bibitem{Sorkin:1997ja} 
  R.~D.~Sorkin,
  %``The statistical mechanics of black hole thermodynamics,''
  gr-qc/9705006.
  %%CITATION = GR-QC/9705006;%%
  %46 citations counted in INSPIRE as of 01 Dec 2015  

%\cite{Page:2013mqa}
\bibitem{Page:2013mqa} 
  D.~N.~Page,
  %``Excluding Black Hole Firewalls with Extreme Cosmic Censorship,''
  JCAP {\bf 1406}, 051 (2014)
  doi:10.1088/1475-7516/2014/06/051
  [arXiv:1306.0562 [hep-th]].
  %%CITATION = doi:10.1088/1475-7516/2014/06/051;%%
  %33 citations counted in INSPIRE as of 01 Dec 2015
  
  %\cite{Page:1985em}
\bibitem{Page:1985em} 
  D.~N.~Page and K.~C.~Phillips,
  %``Selfgravitating Radiation in Anti-de Sitter Space,''
  Gen.\ Rel.\ Grav.\  {\bf 17}, 1029 (1985).
  doi:10.1007/BF00774206
  %%CITATION = doi:10.1007/BF00774206;%%
  %20 citations counted in INSPIRE as of 02 Dec 2015
  
  %\cite{Vaganov:2007at}
\bibitem{Vaganov:2007at} 
  V.~Vaganov,
  %``Self-gravitating radiation in AdS(d),''
  arXiv:0707.0864 [gr-qc].
  %%CITATION = ARXIV:0707.0864;%%
  %14 citations counted in INSPIRE as of 02 Dec 2015
  
  %\cite{Gentle:2011kv}
\bibitem{Gentle:2011kv} 
  S.~A.~Gentle, M.~Rangamani and B.~Withers,
  %``A Soliton Menagerie in AdS,''
  JHEP {\bf 1205}, 106 (2012)
  doi:10.1007/JHEP05(2012)106
  [arXiv:1112.3979 [hep-th]].
  %%CITATION = doi:10.1007/JHEP05(2012)106;%%
  %23 citations counted in INSPIRE as of 02 Dec 2015

%\cite{Pesci:2006sb}
\bibitem{Pesci:2006sb} 
  A.~Pesci,
  %``Entropy of gravitating systems: Scaling laws versus radial profiles,''
  Class.\ Quant.\ Grav.\  {\bf 24}, 2283 (2007)
  doi:10.1088/0264-9381/24/9/009
  [gr-qc/0611103].
  %%CITATION = doi:10.1088/0264-9381/24/9/009;%%
  %7 citations counted in INSPIRE as of 10 Dec 2015
    
%\cite{Anastopoulos:2011av}
\bibitem{Anastopoulos:2011av} 
  C.~Anastopoulos and N.~Savvidou,
  %``Entropy of singularities in self-gravitating radiation,''
  Class.\ Quant.\ Grav.\  {\bf 29}, 025004 (2012)
  [arXiv:1103.3898 [gr-qc]].
  %%CITATION = ARXIV:1103.3898;%%
  %2 citations counted in INSPIRE as of 24 Apr 2015

\bibitem{Zurek}
W. H. Zurek and D. N. Page, Phys.\ Rev.\ D {\bf 29}, 628 (1984).

%\cite{Anastopoulos:2014zqa}
\bibitem{Anastopoulos:2014zqa} 
  C.~Anastopoulos and N.~Savvidou,
  %``The thermodynamics of a black hole in equilibrium implies the breakdown of Einstein equations on a macroscopic near-horizon shell,''
  JHEP {\bf 1601}, 144 (2016)
  doi:10.1007/JHEP01(2016)144
  [arXiv:1410.0788 [gr-qc]].
  %%CITATION = doi:10.1007/JHEP01(2016)144;%%
  %2 citations counted in INSPIRE as of 02 Dec 2016

\end{thebibliography}
\end{document}